\documentclass{jfm}
\usepackage{graphicx}
\usepackage{epstopdf, epsfig}

\usepackage{natbib}
\usepackage{graphicx}
\usepackage{float}
\usepackage{subfigure}
\usepackage{amsmath}
\usepackage{amssymb}
\usepackage{color}
\usepackage{soul}

\graphicspath{{Figures/}}

\newcommand{\ii}[0]{\mathrm{i}}

\newcommand{\ee}[0]{\mathrm{e}}

\shorttitle{A1 and A2 modes in jet screech}
\shortauthor{P. A. S. Nogueira et al.}

\title{Closure mechanism of the A1 and A2 modes in jet screech}

\author{Petr\^onio A. S. Nogueira\aff{1}
  \corresp{\email{petronio.nogueira@monash.edu}},
  Vincent Jaunet\aff{2},
  Matteo Mancinelli\aff{2},
  Peter Jordan\aff{2}
  \and Daniel Edgington-Mitchell \aff{1}}
  
\affiliation{\aff{1}Department of Mechanical and Aerospace Engineering, Laboratory for Turbulence Research in Aerospace and Combustion, Monash University, Clayton, Australia
\aff{2}Département Fluides, Thermique, Combustion, Institut PPrime, CNRS–Université de Poitiers–ENSMA, Poitiers, France}

\begin{document}

\maketitle

\begin{abstract}
This paper explores the screech closure mechanism for different axisymmetric modes in shock-containing jets. While many of the discontinuities in tonal frequency exhibited by screeching jets can be associated with a change in the azimuthal mode, there has to date been no explanation for the existence of multiple axisymmetric modes at different frequencies. This paper provides just such an explanation. As shown in previous works, specific wavenumbers arise from the interaction of waves in the flow with the shocks. This provides new paths for driving upstream-travelling waves that can potentially close the resonance loop. Predictions using locally parallel and spatially periodic linear stability analyses and the wavenumber spectrum of the shock-cell structure suggest that the A1 mode resonance is closed by a wave generated when the Kelvin-Helmholtz mode interacts with the leading wavenumber of the shock-cell structure. The A2 mode is closed by a wave that arises due to interaction between the Kelvin-Helmholtz wave and a secondary wavenumber peak, which arises from the spatial variation of the shock-cell wavelength. The predictions are shown to closely match experimental data, and possible justifications for the dominance of each mode are provided based on the growth rates of the absolute instability.
\end{abstract}

\begin{keywords}
Authors should not enter keywords on the manuscript.
\end{keywords}

\section{Introduction}
\label{sec:intro}

Discrete tones have been observed in shock-containing jets since the 1950s. These are associated with the screech phenomenon, first studied by \cite{powell1953} using schlieren photographs, who suggested that this resonance loop was due to a mechanism involving large-scale structures and upstream-travelling acoustic waves. Such assumptions were used in the development of several resonance models focused on predicting screech frequencies for different jet regimes, and most of them are summarised in recent reviews \citep{RAMAN1998,EdgingtonMitchell2019}. The seminal works of \cite{merle1956frequence,DaviesOldfield} identified that screech tones associated with axisymmetric disturbances could actually be separated into two stages, A1 and A2, related to different acoustic tones. They also showed the existence of B, D and C modes, relating to flapping and helical disturbances respectively. The A1 to A2 mode staging is unique in that no change in azimuthal mode accompanies the change in tonal frequency; other discontinuities in frequency are accompanied (and presumably driven) by a change in the azimuthal mode $m$ of screech, in the case of transition from A to B stages, or by a change in the phase relationship between $m=\pm1$, in the B/D (flapping) to C (helical) stages \citep{EdgingtonMitchell2019}. This property of the A1 and A2 modes has driven efforts to seek alternative explanations for the mechanism behind the frequency change, including different closure mechanisms for the resonance loop \citep{ShenTam2002}. However, recent studies have shown that, for both the axisymmetric A1 and A2 screech modes, the resonance phenomenon is underpinned by the downstream-travelling Kelvin-Helmholtz wavepacket and guided, upstream-travelling jet modes \citep{edgington-mitchell_jaunet_jordan_towne_soria_honnery_2018,gojon2018aiaa}; the change in frequency cannot be explained by a change in the nature of the upstream-propagating wave. The latter belongs to a branch of discrete modes of the stability eigenspectrum associated with a waveguide behaviour of the jet \citep{tam_hu_1989}, and only becomes discrete at specific (cut-on) frequencies, for which their phase velocity is below the sound speed. Prediction models using this upstream-travelling jet mode are in good agreement with experiments \citep{mancinelli2019}, prevailing over models that consider acoustic waves for resonance closure.

Even though waves can be described using models based on the physics of the problem, such as the vortex-sheet formulation, semi-empirical relations are often used to obtain a wavenumber relationship between upstream- and downstream-travelling waves, such that screech frequencies can be predicted. \cite{TAM1986} followed a different approach, and considered screech as special case of broadband shock-associated noise. In this framework, as formulated by \cite{TamTanna1982}, acoustic waves are generated by the interaction between instability waves and the shock-cell structure, which generates sound in directional patterns. Thus, the authors stated that screech could be seen as the limit of this theory when radiation in the upstream direction is considered. Some aspects of this phenomenon were confirmed by \cite{ShenTam2002}, whose model predictions using the shock-cell dominant wavenumber were comparable with experiments for the A1 and B modes, even though acoustic waves were used in their predictions. Still, no clear verification of this mechanism was provided for the A2 mode.

The theory developed by \cite{TamTanna1982} was verified both experimentally and by linear stability models in \cite{EdgingtonMitchell2020}. In this previous work, the authors extracted the most energetic structure associated with the resonance loop by means of a Proper Orthogonal Decomposition (POD) applied to particle image velocimetry of screeching jets. The expected wavenumber relation between the different waves in the flow was obtained from the spatial Fourier spectrum of the travelling wave built from the dominant POD modes. The same wavenumber relation was observed in the most unstable mode from global stability analysis performed around the experimental mean flow, providing a further confirmation that an energy redistribution mechanism is at play in shock-containing jets. The recent work of \cite{nogueira2021splsa} shed light on this underlying energy redistribution mechanism. In that work, the authors showed that screech is actually caused by an absolute instability mechanism induced by the spatial periodicity of the flow, imposed by the presence of a shock-cell structure. Comparison between screech frequencies and shapes of the modes close to the saddle-point in the spatial spectrum are in line with recent experiments, confirming that such instability is the source of the sharp tones observed in the far-field. The main parameters that define the frequency of the saddle-point in the complex plane are the wavenumbers of the KH and guided jet modes, and the shock-cell wavenumber.

The main focus of this paper is to provide a clarification on the underlying mechanisms responsible for the generation of A1 and A2 screech modes in a supersonic imperfectly-expanded jet. For that, two methods are proposed. First, we study screech generation by analysing the different waves that the flow can support at several streamwise stations. Instead of using semi-empirical relations, or considering a given shock as a reflection point for an incident KH wave \citep{mancinelli2019}, we consider that the upstream-travelling waves are generated by interaction between the Kelvin-Helmholtz wavepacket and the shock-cell structure, as in \cite{TamTanna1982}. We then analyse the frequency of the saddle-points in the complex plane via spatially periodic linear stability analysis using an analytical flow model. We start by presenting the modelling methods in \S\ref{sec:method}. In \S\ref{sec:experiments} we describe the experimental setup for the evaluation of mean flows and sound spectra as a function of ideally expanded Mach number $M_j$, and in \S\ref{sec:pressureshocks} we show some key characteristics of the shock-cell structure that will distinguish the mechanisms of A1 and A2 screeching modes. Results of the modelling are shown in \S\ref{sec:resultsA1A2}, where the dominance of either screech mode is obtained by analysing the spatio-temporal growth rate of the absolute instability. The paper is closed with conclusions in \S\ref{sec:concl}.

\section{Screech-frequency evaluation methods}
\label{sec:method}

\subsection{Locally parallel linear stability analysis}
\label{sec:LSAmethod}

The first method is based on the shock-cell structure in supersonic jets and on the different waves supported by the flow at each streamwise station. Following the derivation of \cite{TamTanna1982,ShenTam2002}, detailed in \cite{EdgingtonMitchell2020}, the interaction between a Kelvin-Helmholtz wavepacket (peak wavenumber $k_{kh}$) and the shock-cell structure (peak wavenumber $k_{sh}$) transfers energy to specific wavenumbers $k_x$ given by

\begin{equation}
    k_x = k_{kh} \pm k_{sh}.
    \label{eqn:UshocksLin2}
\end{equation}

\begin{figure}
\centering
\subfigure[Locally parallel]{\includegraphics[clip=true, trim= 0 0 0 0, width=0.5\textwidth]{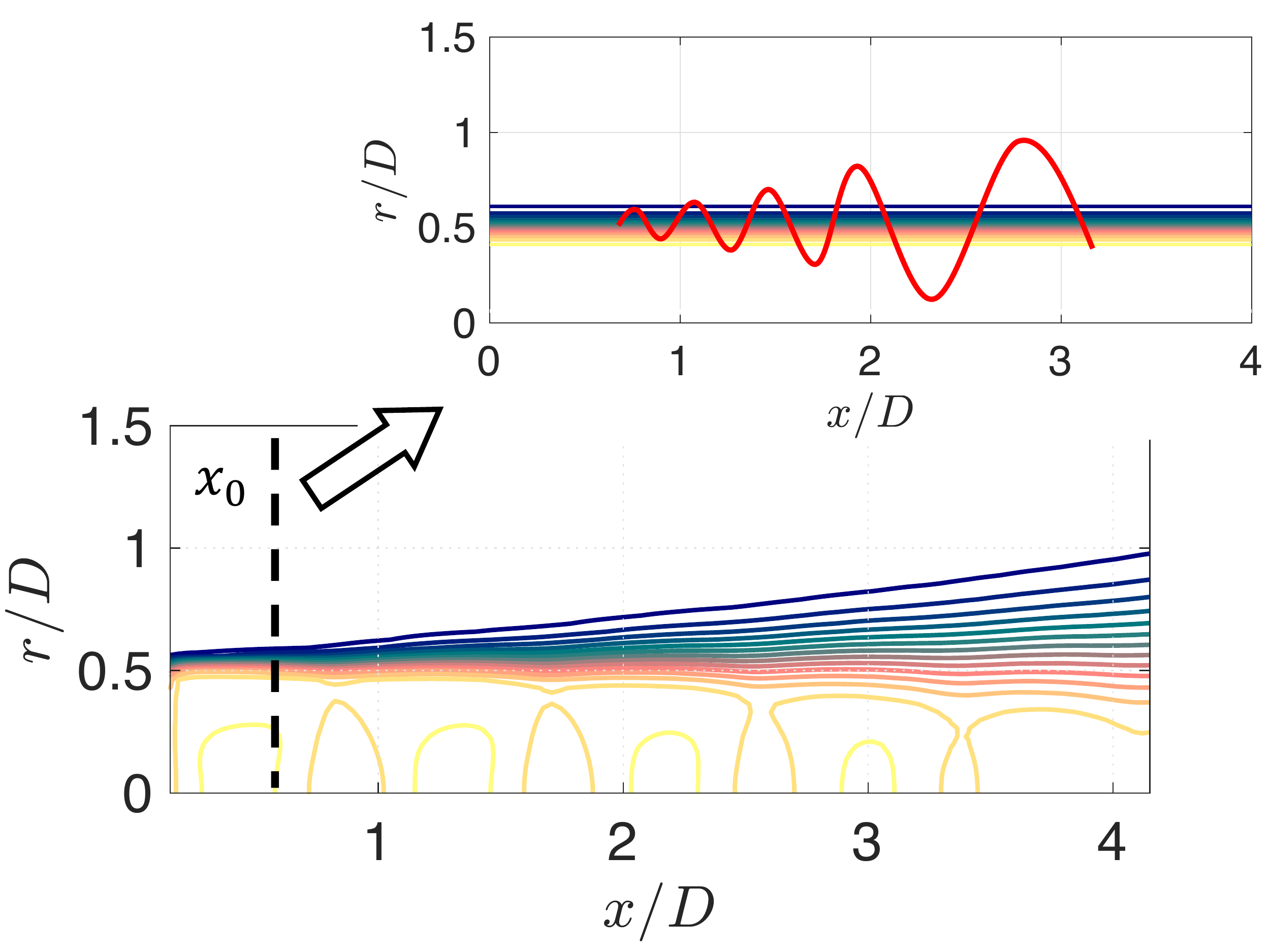}}\subfigure[Spatially periodic]{\includegraphics[clip=true, trim= 0 0 0 0, width=0.5\textwidth]{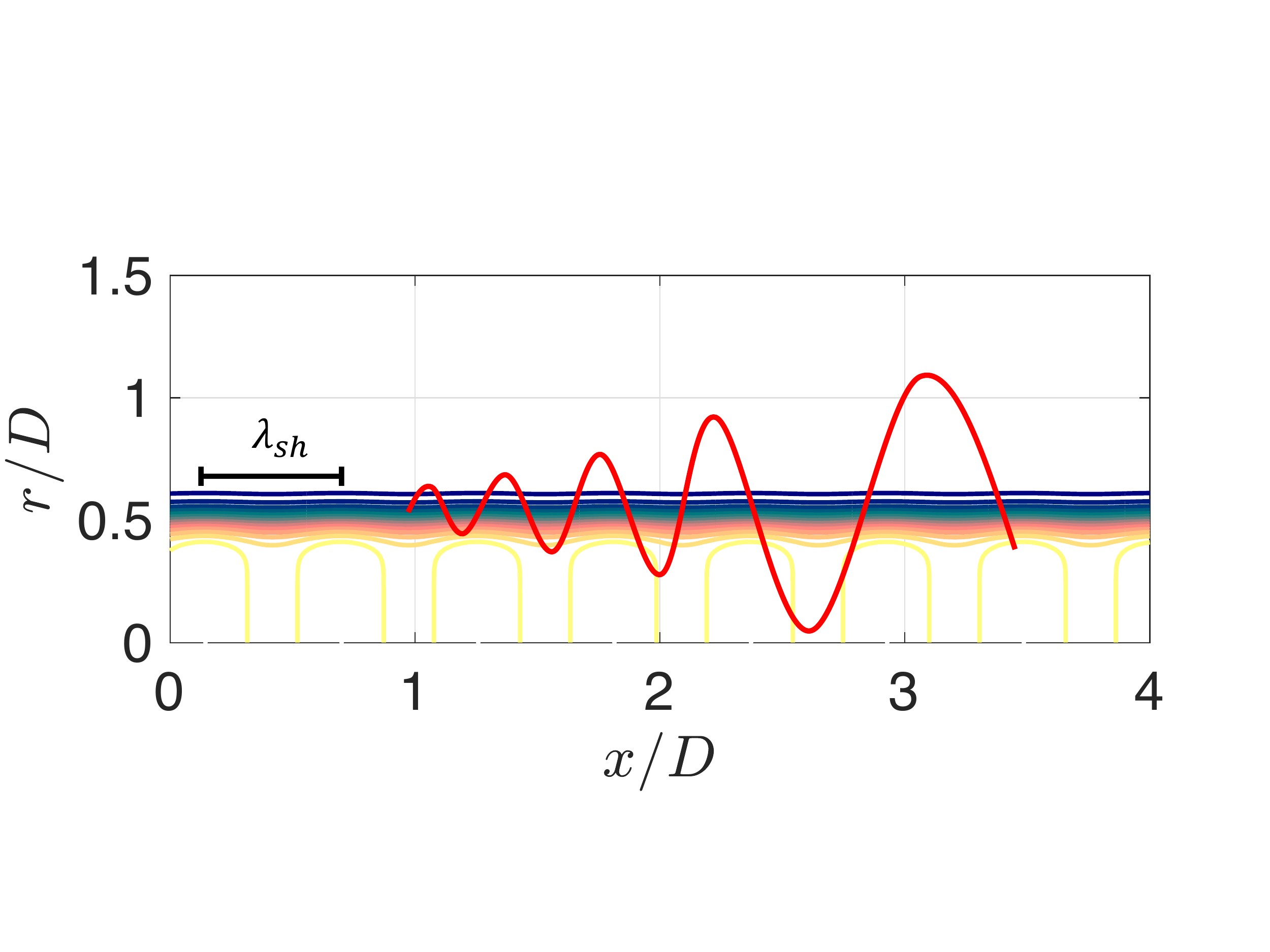}}
\caption{Sketch showing how each wave is computed in the models presented herein. Locally parallel linear stability analysis around a turbulent mean flow (a) and spatially periodic linear stability analysis (b).}
\label{fig:Sketch}
\end{figure}

We here consider characterisation of the A1 and A2 modes based on the streamwise evolution of the shock-cell structure and on spatial linear stability analysis around an experimental mean flow. Only the leading shock-cell wavenumber was considered in previous studies, but it is well known that the shock spacing is a function of streamwise position (as shown in \cite{HARPERBOURNE974}, see also \cite{tam_jackson_seiner_1985,ray_lele_2007}). As will be seen in section \ref{sec:pressureshocks}, a key element of this work is the inclusion of variation in shock spacing. This variation manifests as secondary peaks in the spectral domain, and these secondary peaks are likewise potential sources of interaction as per equation \ref{eqn:UshocksLin2}. The inviscid linearised Navier-Stokes equation can be written using matrix operators in a generalised eigenvalue problem form as

\begin{equation}
    \mathrm{\mathbf{A}} \mathrm{\mathbf{q}} = k_x \mathrm{\mathbf{B}} \mathrm{\mathbf{q}},
   \label{eqn:RayleighEigen}
\end{equation}

\noindent where the disturbance vector $\mathrm{\mathbf{q}}(r)=[ \nu \ u_x \ u_r \ u_\theta \ p]^\mathrm{T}$ includes specific volume, streamwise, radial and azimuthal velocities, and pressure. All variables are considered to have a $\ee^{-\ii \omega t + \ii k_x x + \ii m \theta}$ implicit dependency in this locally parallel framework, and the operators $\mathrm{\mathbf{A}}$, and $\mathrm{\mathbf{B}}$ are dependent on the mean quantities at a fixed streamwise station $x_0$ $\bar{\mathrm{\mathbf{q}}}(x_0,r)=[\bar{\nu} \ U_x \ U_r \ U_\theta \ P]$, on the azimuthal wavenumber, and on the radial derivatives. The influence of the mean radial and azimuthal velocities is considered to be negligible for the low Mach numbers analysed herein ($U_r=U_\theta=0$), and all mean-flow quantities are extracted from experiments, as detailed in section \ref{sec:experiments}. All variables are normalised using the jet diameter $D$, the ambient sound speed $c_\infty$, and the ambient density $\rho_\infty$. Equation \ref{eqn:RayleighEigen} is solved numerically in MATLAB using a Chebyshev polynomials discretisation \citep{trefethen2000spectral} in the radial direction, with $N_r=250$ points. The mapping developed by \cite{lesshafft2007linear} was used to obtain a higher node density in the shear and core regions of the jet, and boundary conditions were implemented as in the cited work. A sketch of the present method is shown in figure \ref{fig:Sketch}(a).

In this framework, equation \ref{eqn:UshocksLin2} fixes a wavenumber relationship between interacting waves in the flow. The Kelvin-Helmholtz wavepacket is the most amplified coherent structure in these jets. Upstream-travelling waves with wavenumber $k_{kh}-k_{sh}$ are thus likely to arise with high amplitudes, which makes them a natural candidate to close the resonance phenomenon in these screeching jets. In order to evaluate the validity of this hypothesis, a three-step analysis is proposed:

\begin{enumerate}
    \item Compute the peak wavenumbers of the shock-cell structure from the mean velocity or pressure fields for each Mach number.
    \item Use spatial linear stability analysis around the mean velocity field close to the nozzle to extract the wavenumber of the Kelvin-Helmholtz mode ($k_{kh}$) as a function of frequency. The same analysis provides the wavenumbers of the upstream travelling waves (both neutral and stable), highlighting all $k^-$ waves supported by the flow.
    \item Compute the intersection between the interaction wave ($k_{kh}-k_{sh}$) and the branch of upstream-travelling waves supported by the flow, obtained from the eigenspectrum computation at several frequencies. This intersection will provide an estimate of the screech frequency for each Mach number analysed.
\end{enumerate}

While the model described in this section may be useful to predict screech tones when a single mode is at play, it provides no argument for the dominance of different tones when there are multiple competing mechanisms. This is addressed in the next section, where the effect of shock-cell periodicity is explicitly explored.

\subsection{Spatially periodic linear stability analysis}
\label{sec:SPLSAmethod}

In the model described in the previous section, the periodicity of the flow is considered \textit{a posteriori}; all wavenumbers are computed assuming a parallel flow, and the energy transfer due to periodicity is imposed by the resonance condition (as per equation \ref{eqn:UshocksLin2}). The model presented in this section considers a periodic flow directly in the stability analysis; the system is linearised around a periodic mean flow in the shape

\begin{equation}
    U_x(x,r)=U(r)\left[1+A_{sh} \cos{\left( k_{sh} x \right)}\right],
    \label{eqn:meanflow_x}
\end{equation}

\noindent where $A_{sh}$, $k_{sh}$ are the shock-cell amplitude and wavenumber. The radial profile of the mean streamwise velocity in each axial station is given by

\begin{equation}
    U(r)=M\left[0.5+0.5\mathrm{tanh}\left(0.5\left(\frac{0.5D_j}{r}-\frac{r}{0.5D_j}\right)\frac{1}{\delta}\right)\right),
    \label{eqn:UmeanRad}
\end{equation}

\noindent where $M=U_j/c_{\infty}$ is the Mach number, $U_j$ and $D_j$ are the ideally expanded jet velocity and diameter, and $\delta$ is a parameter that characterises the shear layer thickness (see \cite{michalke1971instabilitat}). This mean flow is periodic by construction, and is considered to be a first approximation of a jet with an embedded shock-cell structure, as shown in figure \ref{fig:Sketch}(b). This mean flow periodicity allows us to use the Floquet ansatz, which considers solutions in the shape

\begin{equation}
    \mathbf{\hat{q}}(x,r) = \mathbf{\tilde{q}}(x,r)\ee^{\ii\mu x}.
    \label{eqn:response_per}
\end{equation}

\noindent where $\mu=\mu_r+\ii\mu_i$ is the Floquet exponent, and $\mathbf{\tilde{q}}(x,r)$ can be represented using a Fourier series. Replacing equation \ref{eqn:response_per} in the linearised Navier-Stokes system, we can rewrite it as an eigenvalue problem, similar to the locally parallel case:

\begin{equation}
    \mathbf{\tilde{A}} \mathbf{\tilde{q}} = \mathbf{\tilde{B}} \mu \mathbf{\tilde{q}}.
    \label{eqn:eigenvalueFloquet}
\end{equation}

The operators $\mathbf{\tilde{A}}$, $\mathbf{\tilde{B}}$ are an extension of the locally parallel operators $\mathbf{{A}}$, $\mathbf{{B}}$ to the periodic case, and they can be found in \cite{nogueira2021splsa}. As in the locally parallel case, the solution of the present eigenvalue problem leads to waves that can be classified as stable, unstable, or neutral, following the Briggs' criterion \citep{Briggs1964,Brevdo1996}; for instance, downstream-travelling waves will be amplified in space if $\mu_i<0$, and damped if $\mu_i>0$. One of the main differences between the spatially periodic linear stability analysis (SPLSA) and the locally parallel linear stability analysis (LSA) is the shape of the solution: now, instead of having a single wavenumber, each eigenmode is allowed to have energy in wavenumbers following $k_r \pm N k_{sh}$, with $N$ an integer. For that reason, all eigenvalues appear periodically in the complex plane, which causes upstream- and downstream-travelling waves to appear in the same region of the eigenvalue spectrum. As shown by \cite{nogueira2021splsa}, these periodicity effects give rise to an absolute instability, where a saddle-point involving the KH and the guided jet mode is observed for complex frequency $\omega_0=\omega_{0r}+\ii \omega_{0i}$. Overall, the frequencies of the saddle-points and the structure of the mode close to the saddle are in good agreement with experiments, indicating that screech is a result of an absolute instability mechanism.

Considering the good agreement between the frequencies of the saddle-points and the screech frequencies provided in the cited work, tracking saddles for different Mach numbers can be used as a screech prediction tool. As a means of prediction, this method has the advantage of being both empirically verified and physically justified; it is based on the underlying mechanism of screech generation. Several possible frameworks can be constructed to support such an approach to screech prediction: for example, one could use the expression from \cite{pack1950} to obtain the main shock-cell wavenumber as function of Mach number, and try to recreate the $St \times M_j$ plots typical of screech tones (which will also be a function of the shear-layer thickness). In the present work, we follow a slightly different path: instead of using \cite{pack1950}, we obtain the most energetic wavenumbers of the shock-cell structure from experiments, in order to study the transition between A1 and A2 modes. By analysing the spatio-temporal gain of the absolute instability $\omega_{0i}$, it is also possible to determine which mode will be dominant at each Mach number.

The present eigenvalue problem is solved in Matlab using the Arnoldi method (\textit{eigs}). The domain is discretised in both radial and streamwise direction by using Chebyshev polynomials and Fourier modes, respectively \citep{weideman2000matlab}. As in the previous section, radial mapping and boundary conditions are implemented as in \cite{lesshafft2007linear}. Considering that the mean flow has an analytical expression, convergence of the relevant modes is achieved by using $N_r \times N_x=80 \times 31$ in most cases.

\section{Experimental methodology}
\label{sec:experiments}

In order to provide mean velocity fields and to evaluate the different screech frequencies of a round jet, an experimental campaign was conducted at the SUCRÉ (SUpersoniC REsonance) jet-noise facility of the \textit{Institut Pprime} in Poitiers. The stagnation temperature at the nozzle inlet was kept constant at $T=295$ K, and jet exit variables are estimated using isentropic flow equations. The jet operates in an under-expanded condition, issuing from a convergent nozzle of diameter $D=0.01$ m. In the present study, the stagnation pressure was varied in order to obtain jets with ideally expanded Mach numbers ranging the interval $M_j=[1,1.3]$, with spacing $\Delta M_j = 0.005$. Acoustic measurements using an azimuthal array of six microphones were performed at the nozzle exit plane and radial distance $r/D=1$. This allowed decomposition of the acoustic field into azimuthal Fourier modes; since we focus on the A1 and A2 modes, only $m=0$ will be analysed. For more details on the facility and the acoustic experiments performed, the reader can refer to \cite{mancinelli2019}.

Particle-image velocimetry was performed in this flow for discrete values of $M_j = 1.080; 1.120; 1.160; 1.220$. The flow was seeded using Ondina oil particles before entering the stagnation chamber, ensuring a sufficient seeding homogeneity. The particles were illuminated by a $2 \times 50$ mJ Nd-YAG laser and the images were recorded with a 4Mpix CCD camera equipped with a Sigma Macro 105mm. The camera provided a field-of-view of approximately $10 D \times 10 D$. The PIV image pairs were acquired at a sampling rate of 7.2Hz with a $\Delta t$ of 1 $\mu$s. For each configuration a total of 10000 image pairs were acquired in order to obtain well-converged statistics. The images were processed using LaVision's Davis 8.0 software using a multipass iterative correlation algorithm \citep{Willert1991,SORIA1996} starting with interrogation area of $64\times 64$ pixels and finishing with $16 \times 16$ pixels. The overlap between neighbouring interrogation windows was set at $50\%$, leading to a resolution of about $2.5$ vectors per millimetre (\textit{i.e.} 25 vectors per jet diameter) in the measured field. At each correlation pass, a peak validation criterion was used: vectors were rejected if the correlation peak ratio was lower than $1.4$. This value was selected as the minimum acceptable value ensuring validation in the potential regions of the flow while rejecting most of the evident erroneous vectors. Outliers were then further detected and replaced using universal outlier detection \citep{westerweel2005universal}. The mean pressure and density fields are estimated from the velocity fields using a Crocco-Busemann approximation based on isentropic relations, and a spatial integration method, as described by \cite{van2007evaluation}. All fields were interpolated onto the radial mesh used in the linear stability analysis at the required streamwise position. Sample mean streamwise velocity fields can be found in \cite{mancinelli2020}.

\section{Spectral characteristics of the shock-cell structure}
\label{sec:pressureshocks}

\begin{figure}
\centering
\subfigure[Mean pressure at the centreline]{\includegraphics[clip=true, trim= 0 0 0 0, width=0.5\textwidth]{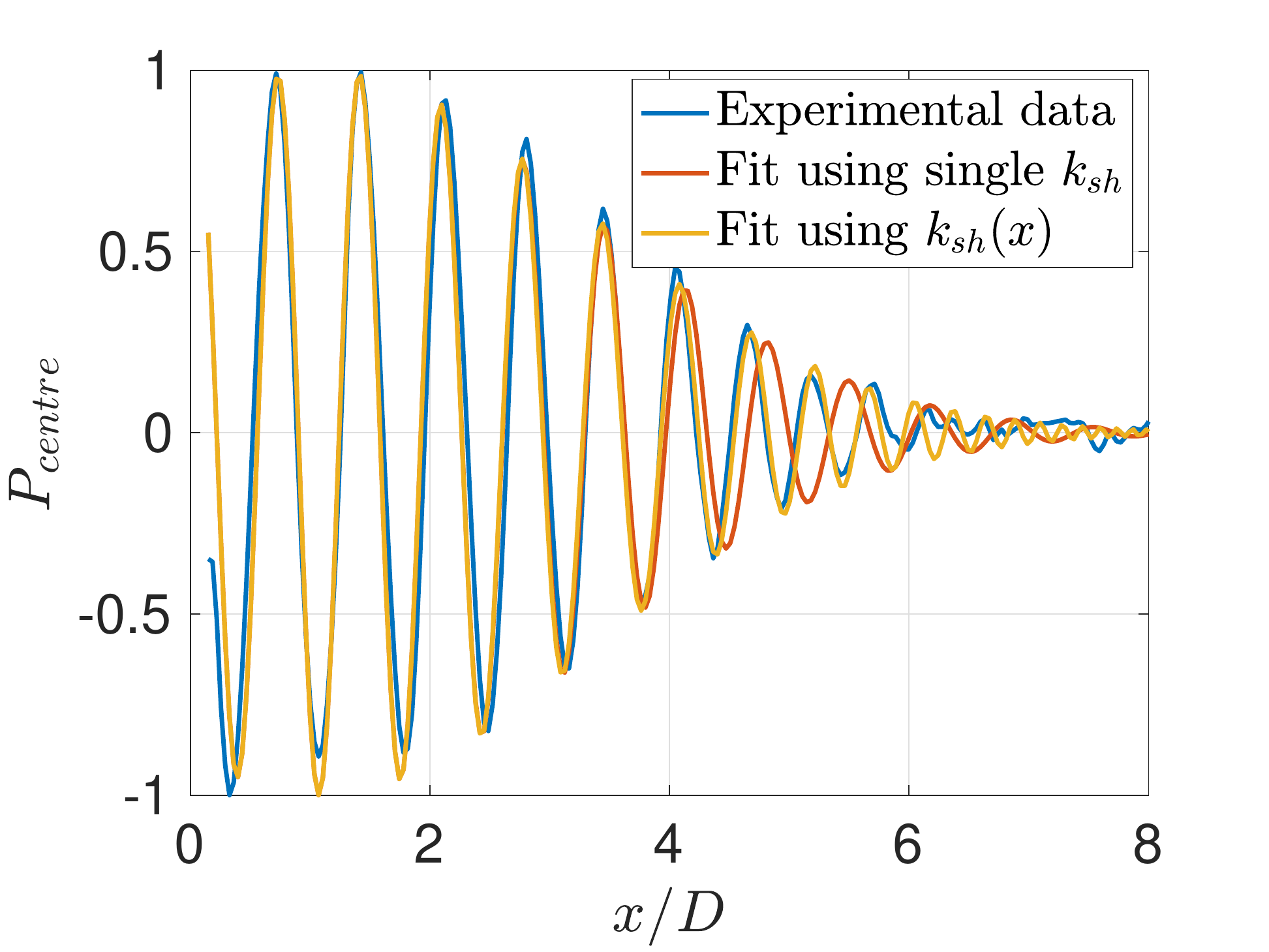}}\subfigure[Spatial Fourier transform of $P_{centre}$]{\includegraphics[clip=true, trim= 0 0 0 0, width=0.5\textwidth]{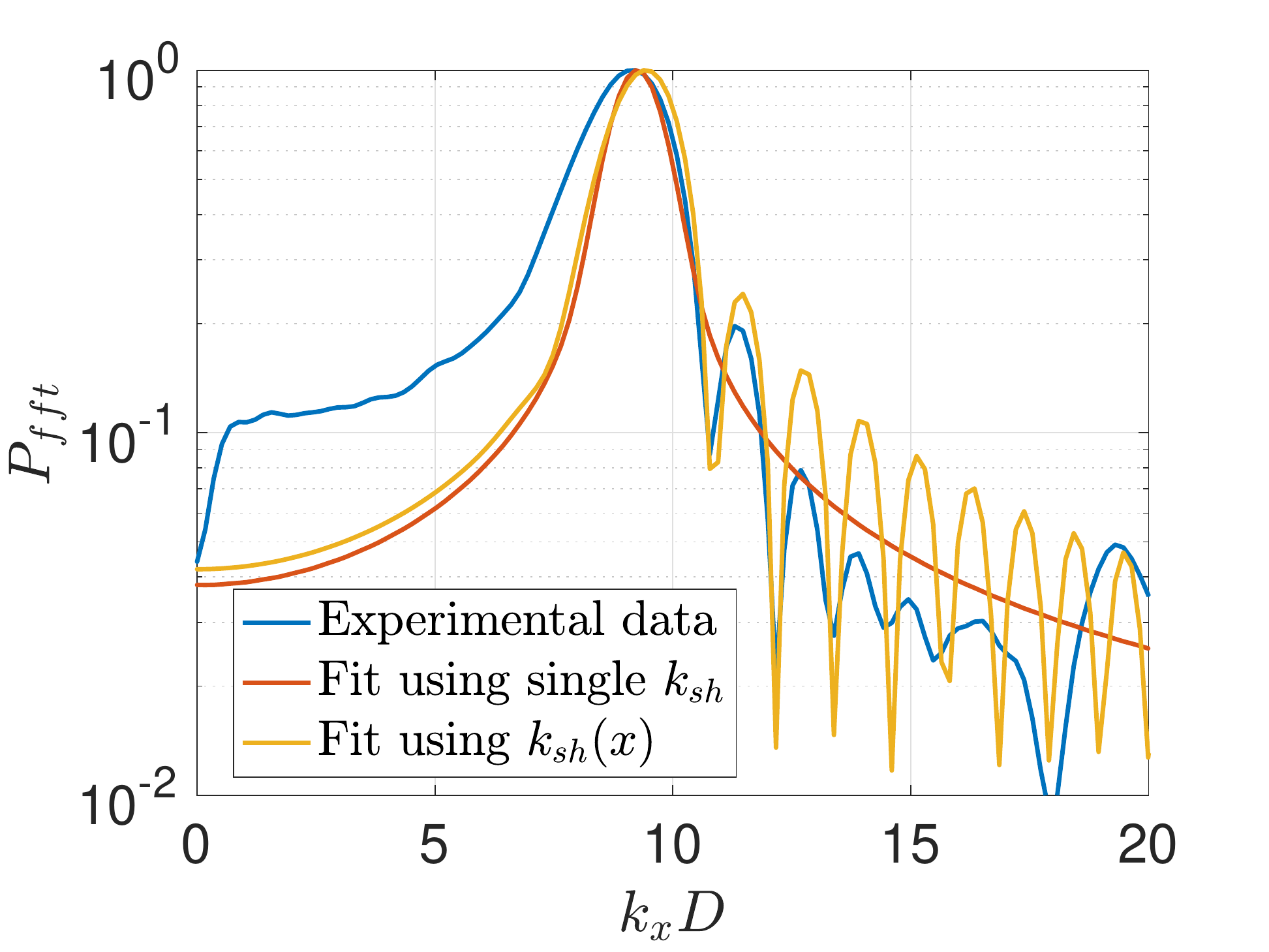}}
\caption{Mean pressure at the centreline educed from data and from different fitting functions for $M_j=1.16$. The streamwise mean was subtracted from $P$ to highlight the oscillatory behaviour. All curves are normalised by their maximum.}
\label{fig:Pfftmodels}
\end{figure}

The relevance of the dominant shock-cell wavenumbers in the redistribution of energy from the wavepacket to the upstream waves motivates an evaluation of the overall spectral characteristics of the shock-cell structure. Figure \ref{fig:Pfftmodels}(a) shows the typical behaviour of $P_{centre}(x)=P(x,0)-\overline{P(x,0)}$ (shown here for $M_j=1.16$), where the overbar denotes the streamwise mean. The distribution resembles a Gaussian-modulated cosine function, displaying an amplitude peak close to the nozzle and a decay further downstream. The streamwise spatial Fourier transform of this field performed across the whole experimental domain is shown in figure \ref{fig:Pfftmodels}(b), where a sharp peak is observed around $k_x D = 10$, followed by secondary peaks at higher wavenumbers. Such behaviour is common in frequency modulated signals, such as chirps and linear time-delayed signals \citep{COOK1967130}, and analytical assessment of such phenomena can be performed using asymptotic methods, such as the method of stationary phase \citep{Murray1984}. 

The spectrum of $P_{centre}$ suggests that the shock-cell structure cannot be represented using a single wavenumber. This is exemplified by fitting two Gaussian-modulated cosine functions to the data: the first has a constant wavenumber $k_{sh}D=9.2138$, and the second has a spatially varying wavenumber $k_{sh}(x)D=9.2138+2.544\times 10^{-5}(x/D)^{6}$. The functions are normalised to have the same maximum amplitude, and their overall shape is given by

\begin{equation}
    P_{fit}(x)=\ee^{-0.1(x/D-1.1)^2} \cos{(k_{sh} x - 0.4608)}.
   \label{eqn:FuntionFit}
\end{equation}

Figure \ref{fig:Pfftmodels}(a) shows that a fit with a single wavenumber correctly follows the experimental data for $x/D<4$; for positions further downstream, the wavelength associated with the shock-cell decreases, in agreement with previous studies \citep{HARPERBOURNE974}. This decrease in wavelength is correctly captured by the spatially varying wavenumber function, which matches the data quite closely throughout the domain. The effect of the spatial variation of $k_{sh}$ is seen in figure \ref{fig:Pfftmodels}(b): while the Gaussian-modulated single-frequency cosine displays a single peak in the spectrum, the streamwise variation leads to the appearance of several secondary peaks in positions that agree well with the experimental data.

\begin{figure}
\centering
\includegraphics[clip=true, trim= 0 0 0 0, width=0.7\textwidth]{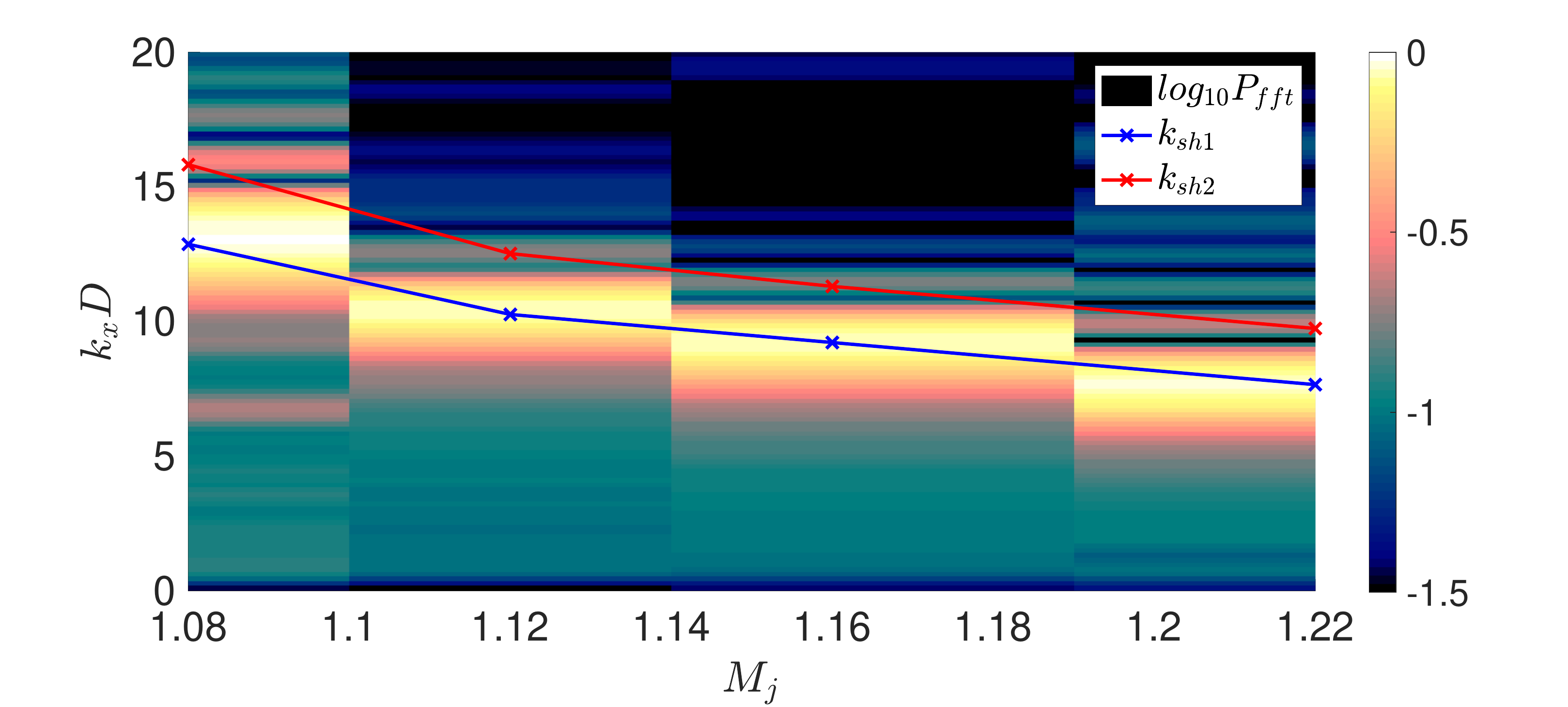}
\caption{Normalised spatial spectrum of the mean pressure field at the centreline as a function of $M_j$. Experimental datapoints are depicted by crosses.}
\label{fig:Pfft}
\end{figure}

The wavenumber spectra for all values of $M_j$ studied in this work are shown in figure \ref{fig:Pfft}, where the presence of secondary peaks is shown to persist throughout the parameter space. The wavenumbers of the two first energetic peaks decay with the increase of $M_j$ (as predicted by \cite{pack1950}), but the difference between them is kept approximately constant. Considering that the spatial variation of the shock-cell wavenumber generates new peaks in the spectrum, both first and second peaks will be used for the analyses detailed in \S\ref{sec:method}. These two wavenumbers will be denoted $k_{sh1}$ and $k_{sh2}$. In the present work, the spatial Fourier transform is obtained without a windowing function, but the addition of a Hanning window does not affect the positions of the primary and secondary shock-cell wavenumbers. A similar behaviour for the shock-cell structure was also observed by \cite{morris_miller_2010}, but the secondary peaks (not associated with the spatial harmonics) were not explored in depth; since these sub-optimal wavenumbers do not agree with the higher-order wavenumbers predicted by \cite{pack1950}, this previous result suggests that those peaks were actually related to the spatial variation of the shock-cell structure.

\section{Results}
\label{sec:resultsA1A2}

In this section we evaluate the performance of both models in predicting the screech tone. One should keep in mind that the models differ in the consideration of periodicity; while the first uses this assumption to obtain an expression for the resonance condition, the second (SPLSA) imposes periodicity directly in the formulation, and resonance is achieved by the presence of a saddle-point between upstream- and downstream-travelling waves with $\omega_{0i}>0$. Still, in order for the saddle to occur, both the KH and the guided jet mode should be close to one another in the complex plane. This means that, close to the saddle, the wavenumber of the upstream wave $k_r^-$ should follow

\begin{equation}
    k_r^- \approx k_{kh} - k_{sh}.
    \label{eqn:UshocksLinSPLSA}
\end{equation}

\noindent for some complex $\omega$. Comparing equations \ref{eqn:UshocksLinSPLSA} and \ref{eqn:UshocksLin2}, it is clear that the first model provides a first approximation of the saddle-point position and real frequency. Thus, even though no information about the spatio-temporal growth rate can be obtained using the first model, both models may still lead to similar screech frequency predictions.

\subsection{Locally parallel stability analysis}

In order to perform the spatial linear stability analysis around the mean flow in the locally parallel framework, a streamwise position must be chosen. Considering that the initial amplification and phase velocity of the Kelvin-Helmholtz (KH) mode is well described by a vortex-sheet model \citep{michalke1971instabilitat}, positions very close to the nozzle are chosen to approximate the frequency-wavenumber behaviour. Results are presented for the position where the streamwise mean velocity is maximum (the first shock-cell, closer to the nozzle), but the results are relatively insensitive to this choice. The choice of streamwise position, $x_{ups}$, for the calculation of the upstream mode is less obvious; a single position may not be representative of the phenomenon, especially if one considers the changes in frequencies of existence of these waves with increasing $x_{ups}$. For that reason, the spectra associated with several mean flow positions with respect to the shock-cell structure were analysed. Such positions were taken as the local maxima and minima of streamwise velocity at the centreline, up to the fourth shock-cell; naturally, these locations are dependent on the flow Mach number, but since they are all connected to the shock-cells, they will provide a fair comparison between the different cases. These positions were used to evaluate the overall frequency-wavenumber characteristics of the upstream wave, an ensemble of tone-frequencies being computed associated with an ensemble of streamwise locations. These positions include the downstream reflection points considered in the prediction model of \cite{mancinelli2019} (defined as the fourth shock-cell) for the whole range of $M_j$. All results are presented using the Strouhal number $St=\frac{\omega D}{2 \pi U_j}$, where $U_j$ is the ideally expanded jet velocity.

\begin{figure}
\centering
\subfigure[$M_j=1.08$]{\includegraphics[clip=true, trim= 0 0 0 0, width=0.5\textwidth]{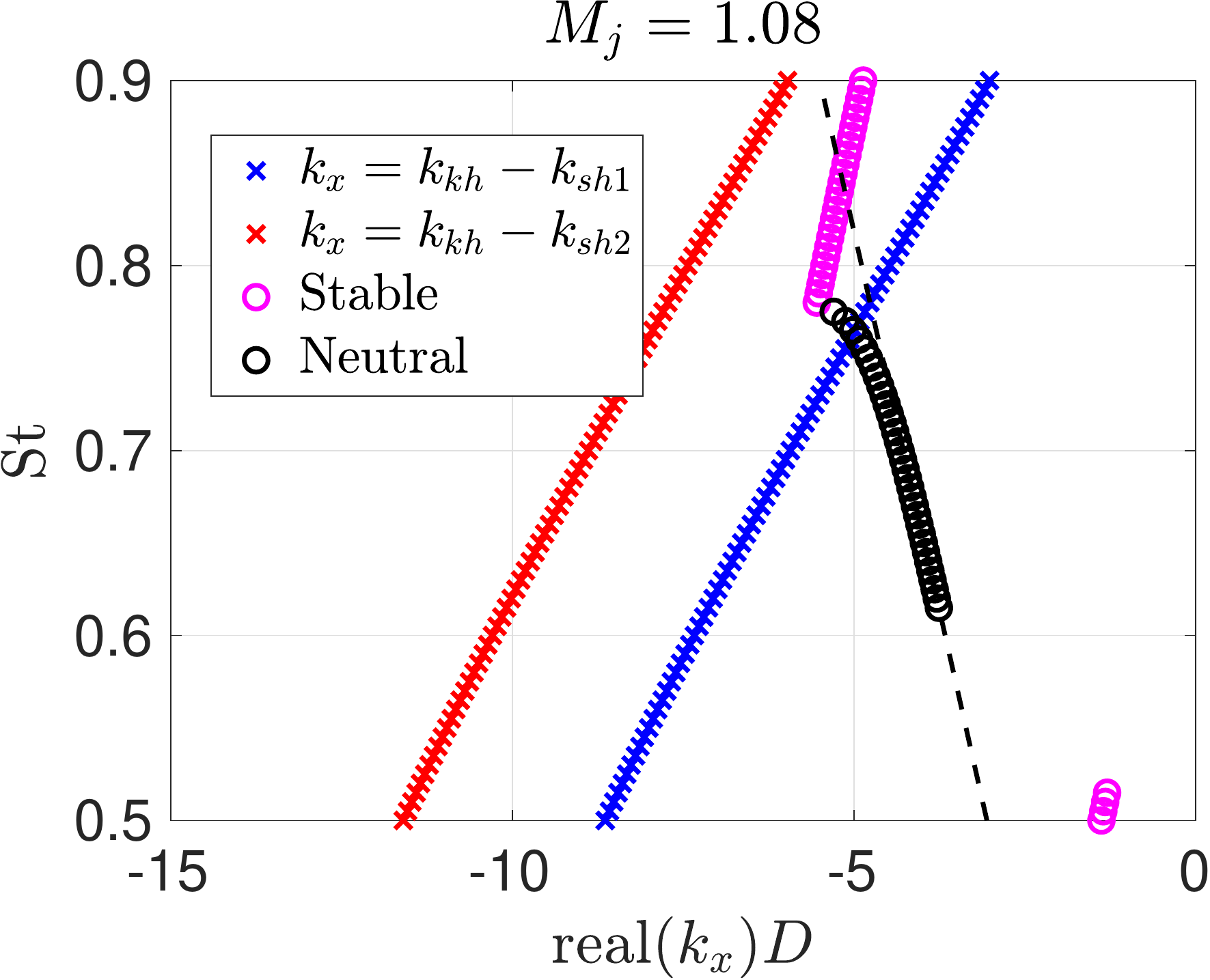}}\subfigure[$M_j=1.16$]{\includegraphics[clip=true, trim= 0 0 0 0, width=0.5\textwidth]{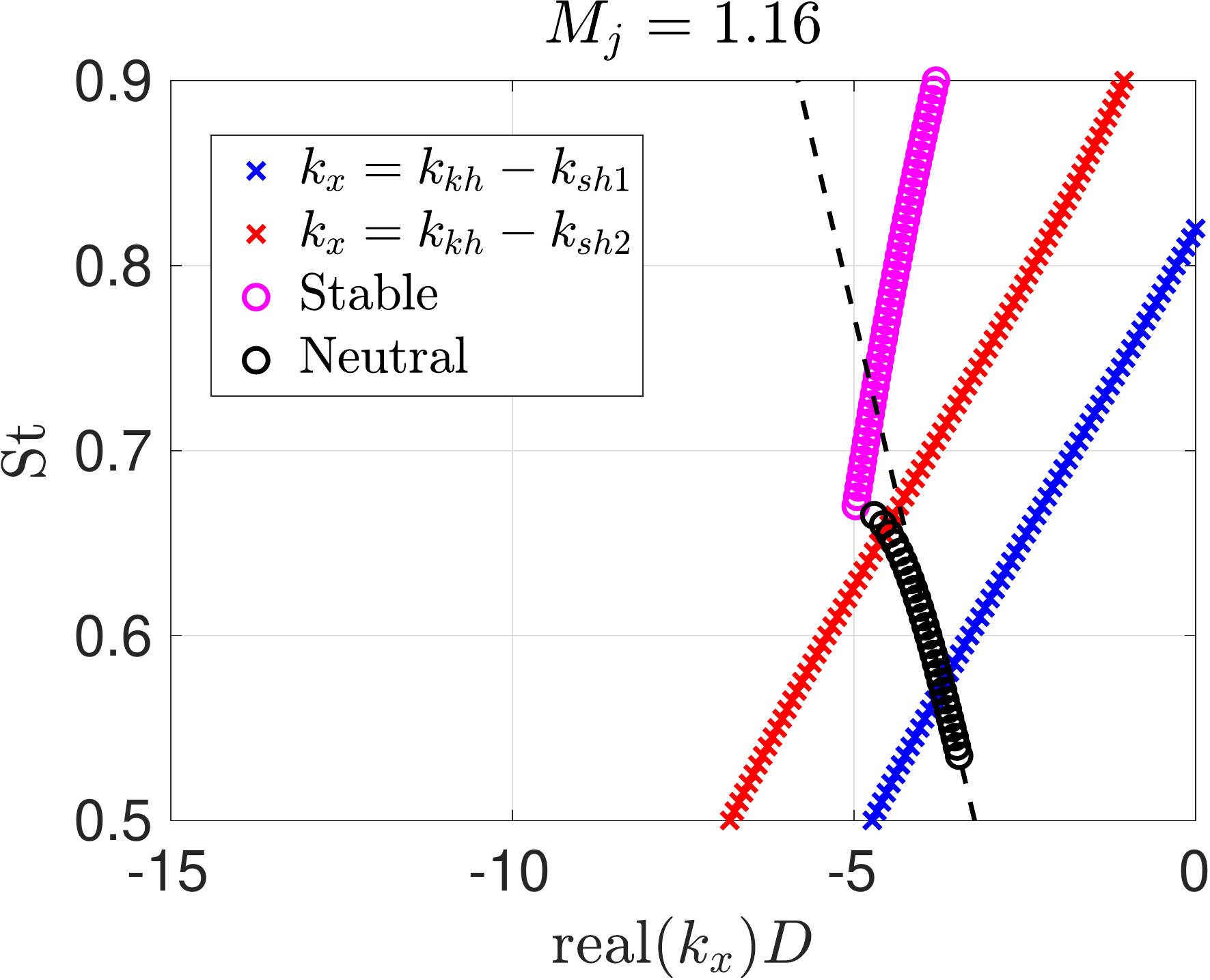}}
\caption{Wavenumbers of the upstream-travelling waves (circles) and wavenumbers forced by the interaction between the Kelvin-Helmholtz mode and the shock-cells (crosses) as function of Strouhal number for $x_{ups}$ considered as the first shock cell position and $M_j=1.08$ (a) and $1.16$ (b). The wavenumber of acoustic waves is represented by the dashed line.}
\label{fig:ResultsMaxCross}
\end{figure}

An illustration of the method described in \S\ref{sec:LSAmethod} is presented in figure \ref{fig:ResultsMaxCross}, where $x_{ups}$ was chosen as the first velocity maximum. In black and pink circles, the dispersion relation of the discrete neutral and stable upstream waves are shown as a function of $St$ for $M_j=1.08$ and $1.16$. These curves are quite similar to those found by \cite{tam_hu_1989, towne_cavalieri_jordan_colonius_schmidt_jaunet_bres_2017}, and the region of existence of the neutral modes is roughly in agreement with that provided by a vortex-sheet model \citep{edgington-mitchell_jaunet_jordan_towne_soria_honnery_2018,gojon2018aiaa,mancinelli2019}, but the cut-on frequency (the lowest frequency in which the guided jet mode exists) is slightly lower. The stable modes are also shown (in pink) here for completeness, as resonance involving this mode may be possible if it is weakly damped in space (small $|imag(k_x)|$) \citep{towne_cavalieri_jordan_colonius_schmidt_jaunet_bres_2017}; in the present analysis, the overwhelming majority of the tones predicted were related to a resonance that includes the neutral modes. The real part of the wavenumbers energised by the interaction of the unstable Kelvin-Helmholtz mode and the shock-cells, i.e. $real(k_{kh} - k_{sh1})$ (blue) and $real(k_{kh} - k_{sh2})$ (red) are also shown. As shown in figure \ref{fig:ResultsMaxCross}(a), the blue symbols intersect the upstream branch at a single frequency; upstream-travelling waves may thus arise as a result of the interaction. For this value of $M_j$, no intersection is found for the red curve, related to the interaction with second peak of the wavenumber spectrum. This implies that at this condition, interaction wavenumbers that arise when the KH wave interacts with the secondary shock cell mode are not matched by a propagative mode. The jet is therefore unable to guide this wavenumber interaction upstream, and resonance does not occur. The effect of varying $M_j$ is shown in figure \ref{fig:ResultsMaxCross}(b), where $M_j=1.16$ was chosen. For this case, the highest and lowest frequencies of existence of the neutral mode occur at lower frequencies, and the reduction of the wavenumbers of the shock-cell structure leads to two intersections with the upstream branch. Thus, the model suggests that resonance can be be closed at two different frequencies: the first related to an interaction of the wavepacket with $k_{sh1}$, and the second related to an interaction with $k_{sh2}$. It is worth noting that the second interaction occurs close to the maximum frequency of existence of the neutral guided jet mode, hindering the possibility of an interaction with other higher-frequency peaks of the shock-cell structure.

\begin{figure}
\centering
{\includegraphics[clip=true, trim= 0 0 0 0, width=\textwidth]{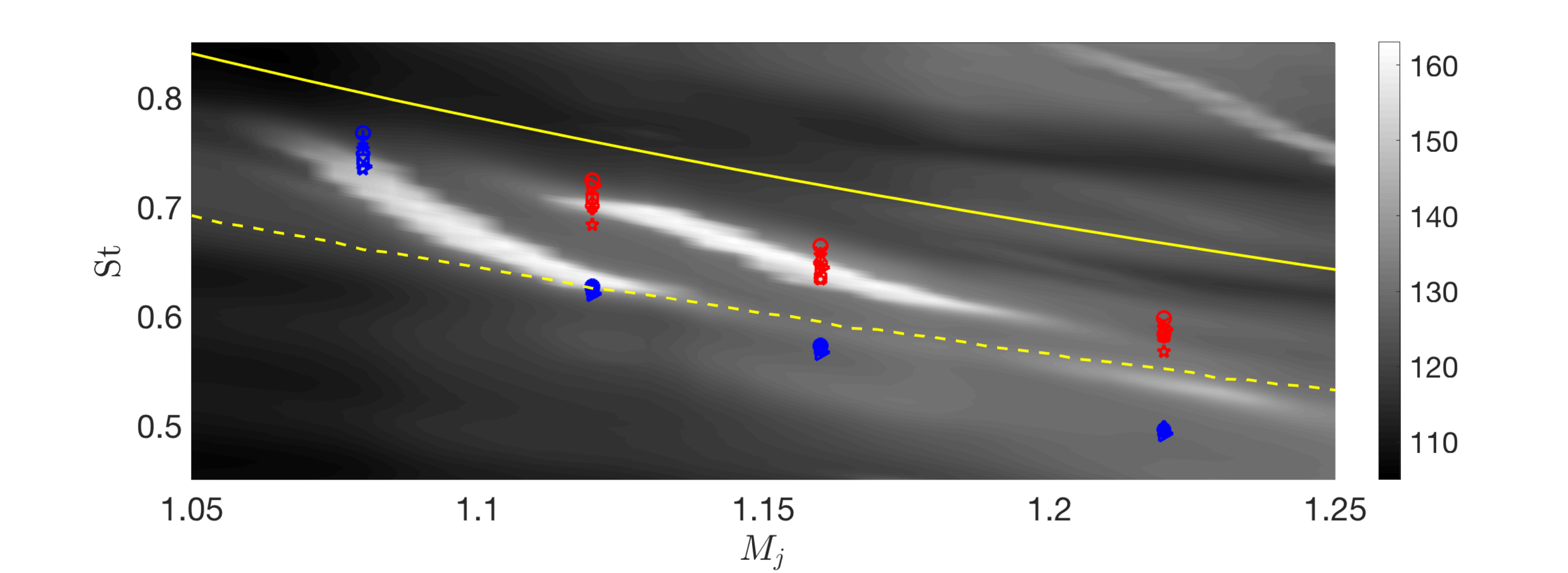}}
\caption{Comparison between the frequencies predicted by the model (symbols) and the PSD map of a screeching jet as function of $M_j$ (a). Prediction of the A1 and A2 modes are depicted by blue and red curves, respectively, and the different $x_{ups}$ chosen as the local maxima and minima of the velocity (local extrema of the shock-cells), up to the fourth shock-cell.}
\label{fig:ResultsMax}
\end{figure}

The comparison between the frequencies predicted by the method and the experimental power spectral density (PSD) for the different $M_j$ and all values of $x_{ups}$ (from the first to the fourth shock-cell) is shown in figure \ref{fig:ResultsMax}. Interaction of the Kelvin-Helmholtz with the main peak of the cell spectrum ($k_{sh1}$), in blue, produces an accurate prediction of the A1 screech frequencies for all cases where this mode is expected to be dominant ($M_j<1.16$), with small variations in the prediction with the increase of $x_{ups}$. As mentioned previously, the A1 mode ceases to exist for higher values of $M_j$ due to a decrease of the cut-on frequency of the discrete guided jet wave with increase of $M_j$. A similar trend is obtained for the A2 mode (red): for $M_j \geq 1.12$, where this mode dominates, the model leads to frequencies quite close to the peaks in the experiments, with slight variations as $x_{ups}$ increases. Such variations in both modes are mostly identified when predictions are close to the maximum frequency of the neutral mode, which may be connected to the steeper changes in the group velocity of such a wave. No intersection is found for lower values of $M_j$ for any of the values of $x_{ups}$ analysed here, which highlights that this mode can only exist for higher Mach number, as expected for the A2 resonance. In accordance to the observations of \cite{mercier_castelain_bailly_2017}, who stated that the upstream wave has an effective source around the fourth shock-cell, it is shown that such wave is supported by the flow in all stations of the flow from this position up to the nozzle for the A2 resonance. The same is true for the A1 case, except at the highest value of $M_j$, where no intersection was found for $x_{ups}$ chosen as the first minimum of the shock-cell structure, close to the nozzle; this suggests that this mode may not be able to return to the nozzle and close resonance at such high Mach number. Also, considering that the appearance of a secondary peak in figure \ref{fig:Pfftmodels}(b) is due to the variation of the shock-cell spacing further downstream, the two resonance cycles can be associated with different regions of the jet in the present model: A1 is related to dynamics closer to the nozzle, while A2 is related to the downstream region, which is also in line with the results of \cite{mercier_castelain_bailly_2017}. The good agreement between the present model and the experimental screech tones for different $x_{ups}$ (which induces only slight variations) underlines the robustness of the analysis. Figure \ref{fig:ResultsMax} also shows the bounds of existence of neutral waves from \cite{mancinelli2020}, obtained from a vortex-sheet (VS) model; as expected, most frequencies predicted by the present model are inside the region of existence of the neutral mode from such model. Compared to the VS, the branch points of the guided jet mode occur for lower frequencies in the present analysis due to the inclusion of a finite shear-layer thickness, as also observed in \cite{mancinelli2020}.

Some limitations of the model can also be observed when only wavenumber and frequency are considered. The most striking one is that the model does little to explain the selection of either A1 or A2 at Mach numbers where the flow can support both. The spatially periodic analysis addresses this limitation by considering the growth rates associated with each resonance loop. This analysis is performed in the next section.

\subsection{Spatially periodic stability analysis}

We now turn to the spatially periodic analysis. As described in \cite{nogueira2021splsa}, the periodicity of the spatial spectrum induced by the shock-cell structure causes guided jet and KH modes to be in the same region of the spectrum. If the shock-cell amplitude is non-zero, a saddle-point between these two modes can be formed. In this section, we track the saddle frequencies as function of $M_j$ for all the four cases where experimental results are available. As in \cite{nogueira2021splsa}, the shock-cell strength is kept as $A_{sh}=0.02$ (but results are roughly insensitive to this parameter, which also suggests that both $k_{sh1}$ and $k_{sh2}$ may be able to close the resonance loop). In order to evaluate the sensitivity of these results to the shear-layer thickness, $\delta$ was chosen as $0.15$, $0.175$ and $0.20$ for all values of $M_j$. All saddles were computed using the method proposed by \cite{monkewitz_1988}.

\begin{figure}
\centering
\subfigure[A1]{\includegraphics[clip=true, trim= 0 0 0 0, width=0.5\textwidth]{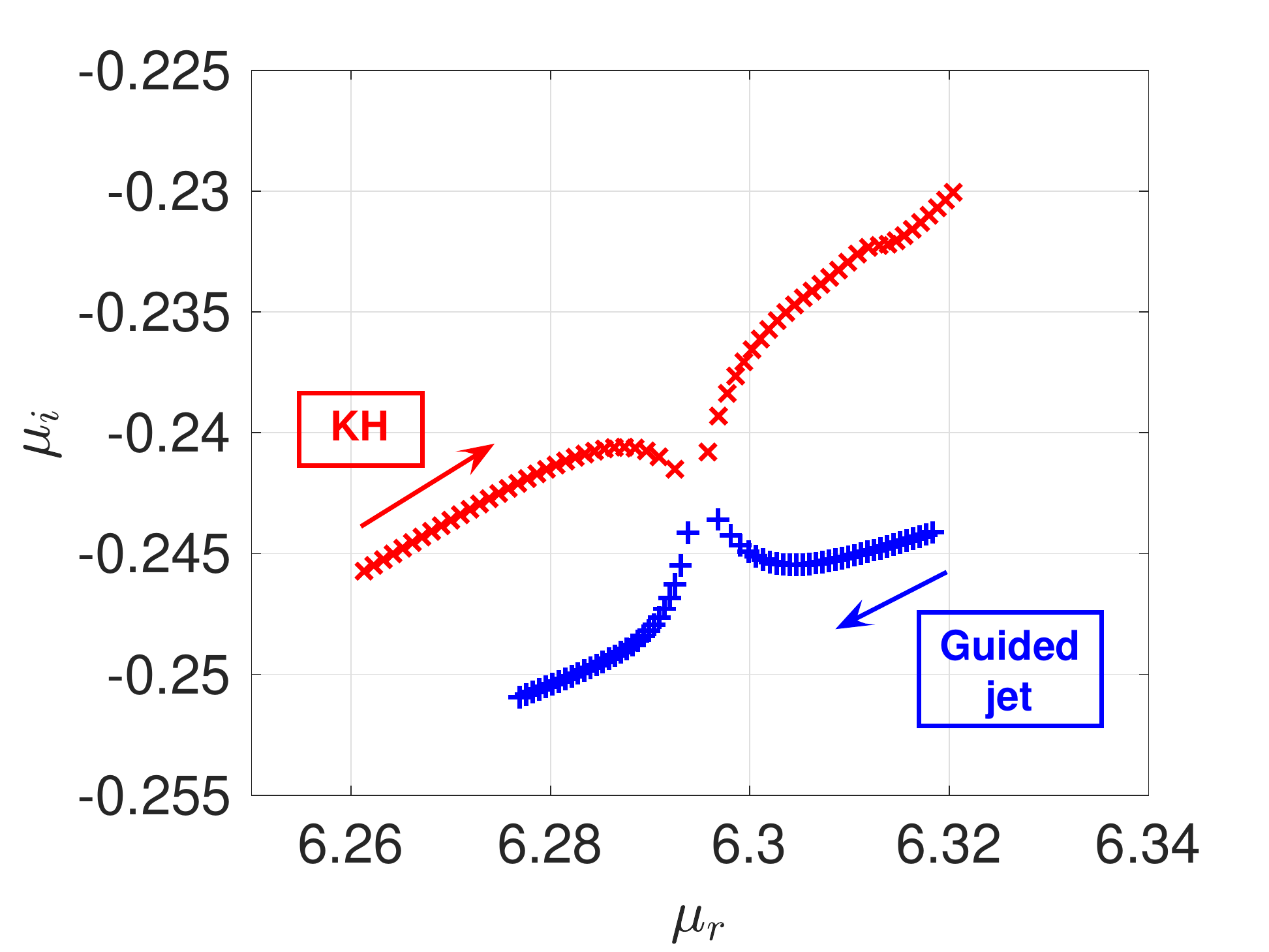}}\subfigure[A2]{\includegraphics[clip=true, trim= 0 0 0 0, width=0.5\textwidth]{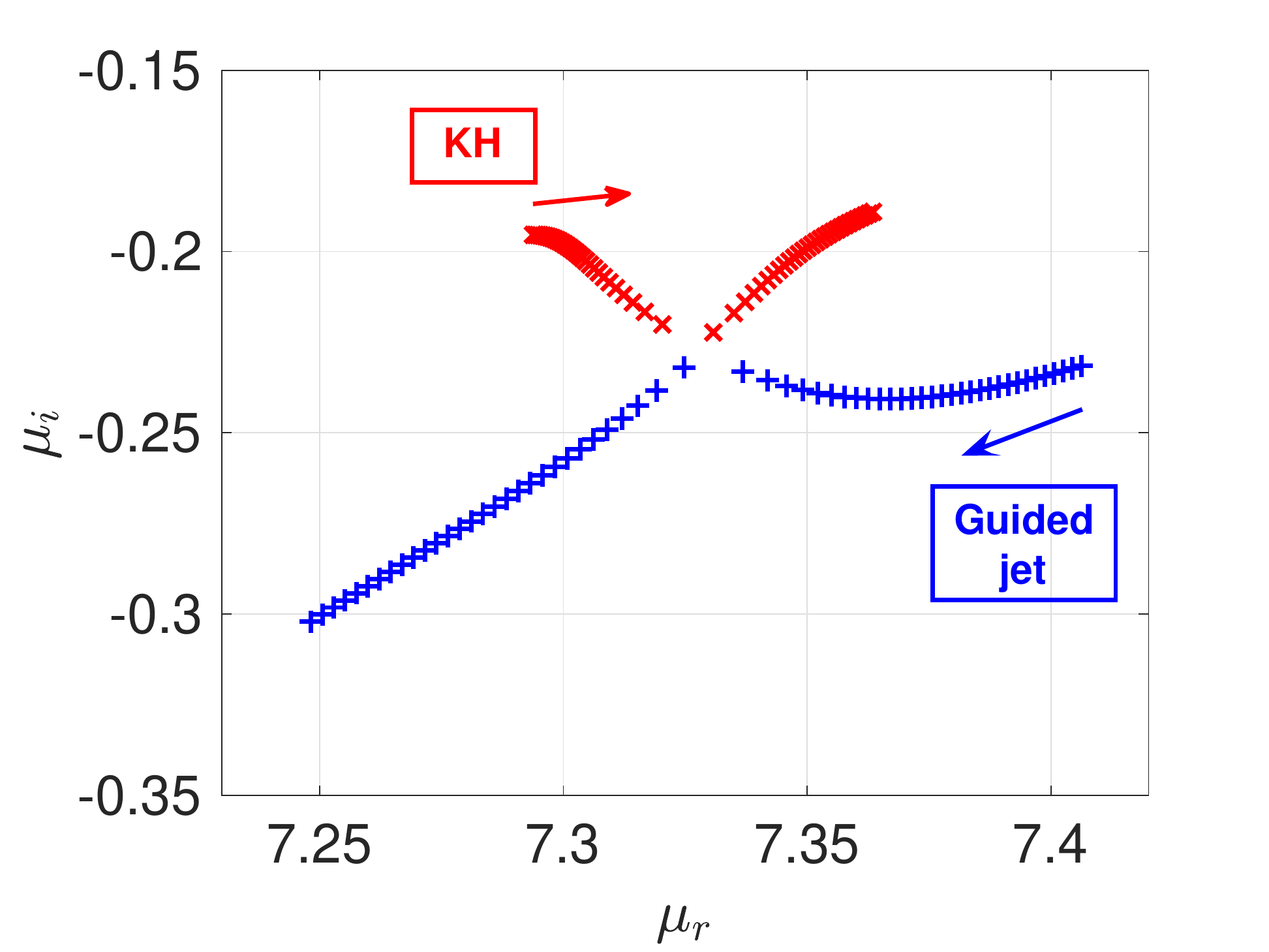}}
\caption{Eigenspectrum of SPLSA in the close to the saddle point for $M_j=1.12$ and $\delta=0.2$. Modes for $k_{sh}=k_{sh1}$, $\omega_{0i}=0.223$ and $0.627<St<0.633$ are shown in (a), and for $k_{sh}=k_{sh2}$, $\omega_{0i}=0.057$ and $0.733<St<0.739$ are shown in (b). Arrows indicate the direction in which each mode travels in the eigenspectrum for increasing $St$.}
\label{fig:SaddlesA1A2}
\end{figure}

In order to exemplify the phenomenon at play, figure \ref{fig:SaddlesA1A2} shows the eigenspectrum close to the saddle point for several Strouhal numbers. This is done for the imaginary frequency of the most unstable saddle, $M_j=1.12$ and $\delta=0.20$. In \ref{fig:SaddlesA1A2}(a), the first shock-cell wavenumber ($k_{sh1}$) is used, and $k_{sh2}$ is used in \ref{fig:SaddlesA1A2}(b). In both cases, Kelvin-Helmholtz modes (indicated in red) travel from left to right, while guided jet modes (indicated in blue) travel from right to left, and the acoustic modes were removed from the spectrum for clarity. These plots show that the trajectory of one mode is modified by the presence of the other in such a way that both are attracted to a single point in the spectrum; after reaching that point, the modes are repelled away, eventually returning to their original trajectories as the distance from the saddle is increased. Following \cite{Brevdo1996}, the resulting double root at the saddle point will grow both downstream and upstream, causing the jet to behave as an oscillator, and generating far-field tones.

\begin{figure}
\centering
{\includegraphics[clip=true, trim= 0 0 0 0, width=\textwidth]{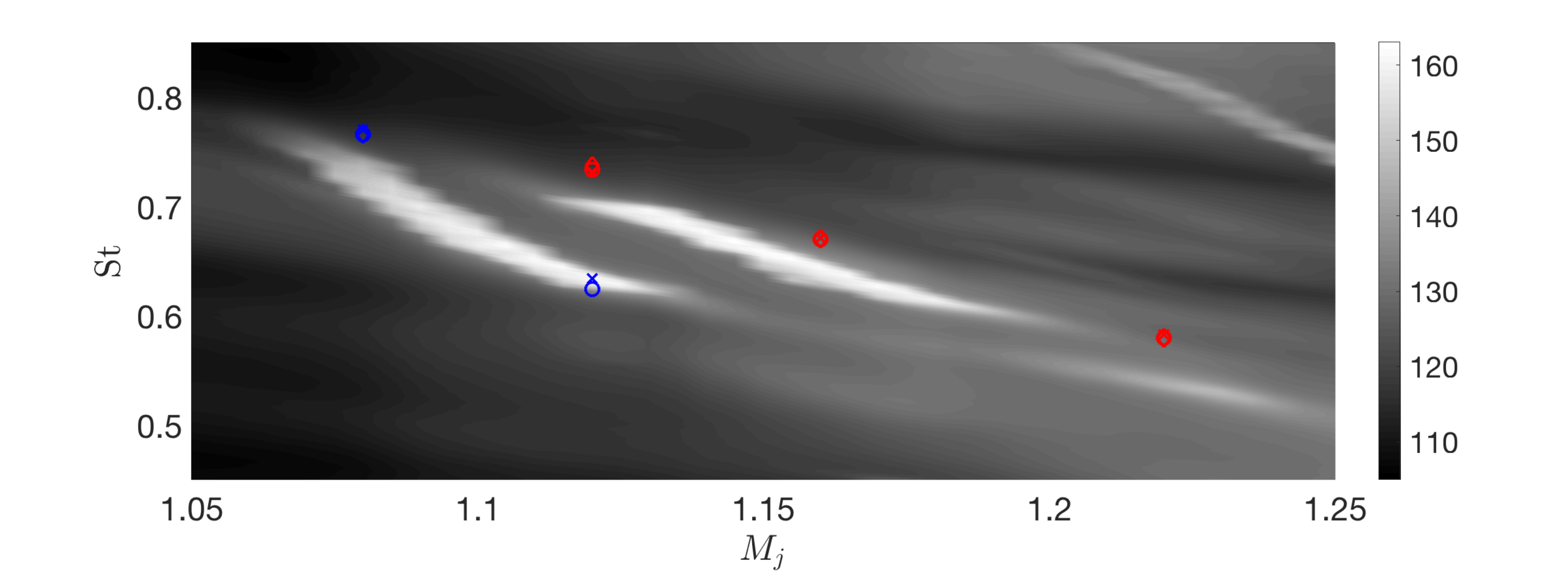}}
\caption{Comparison between the frequencies of the saddle-points from SPLSA (symbols) and the PSD map of a screeching jet as function of $M_j$. Symbols are for $\delta=0.15$ ($\triangle$), $\delta=0.175$ ($\times$) and $\delta=0.20$ ($\circ$).}
\label{fig:PredPSDSPLSA}
\end{figure}

The same process is carried out for the other values of $M_j$ and $\delta$, and the Strouhal number of the saddles are shown in figure \ref{fig:PredPSDSPLSA}. Overall, the predictions align well with the tones observed in the acoustic field, with errors of less than $10\%$ in Strouhal number. Deviations from the predicted values are expected for some reasons: firstly, the exact screech frequency is facility dependent, to some extent (see, for example \cite{gojon2018aiaa}), and deviations of this magnitude could be expected if results from different laboratories are compared. As the present prediction is only based on the wavenumber of the shocks (which is roughly insensitive to the external contours of the nozzle and other details of the facility, as it comes from the solution of \cite{pack1950}), it could be compared to any facility; thus, it may be considered as a reference value for this case. Secondly, it is likely that the actual equivalent ideally expanded jet profile is different from expression \ref{eqn:UmeanRad}, which could change the predictions slightly. Furthermore, even though SPLSA includes a surrogate for the shock-cell in the model, it does not account for effects such as the jet spreading and the streamwise decay of the shocks; these effects might also be at play in such a way to decrease the frequency associated with the resonance phenomenon. Considering the simplicity of the model, that level of agreement for both A1 and A2 tones is still rather remarkable.

\begin{figure}
\centering
\includegraphics[clip=true, trim= 0 0 0 0, width=\textwidth]{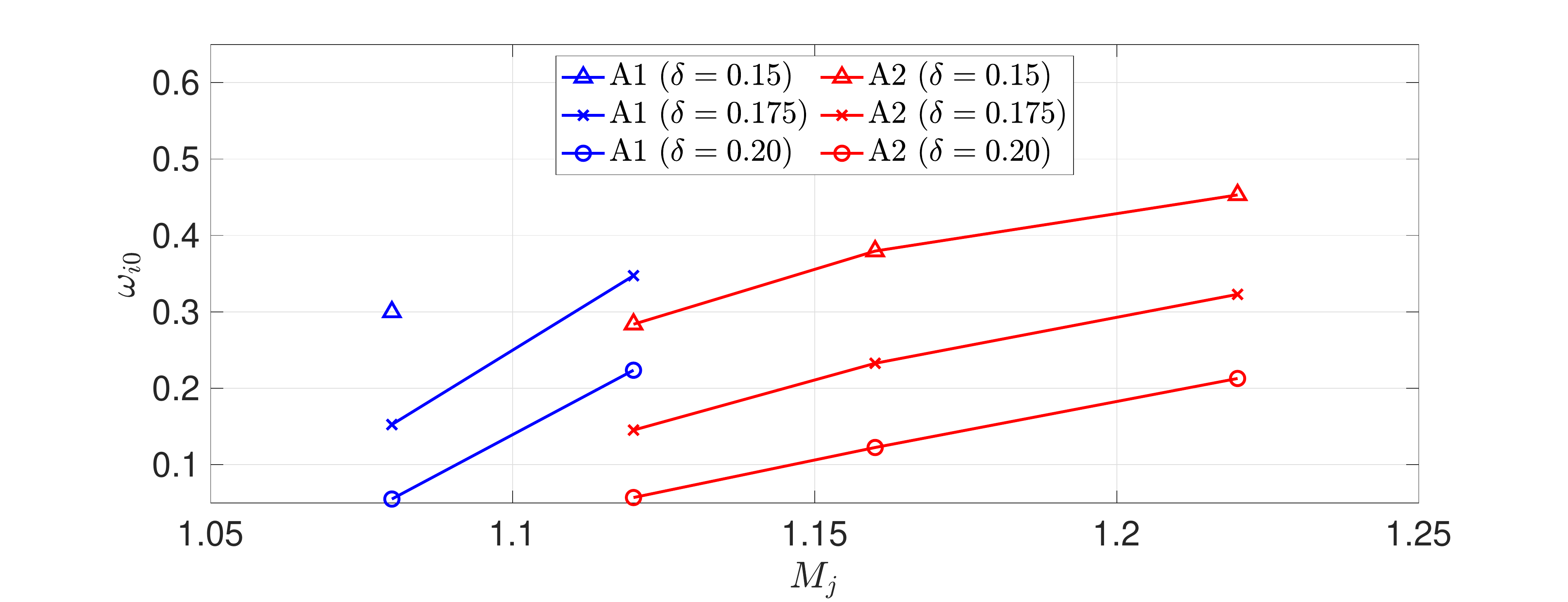}
\caption{Imaginary part of the saddle-point frequency as function of $M_j$ for several values of $\delta$. Both A1 and A2 saddles are shown.}
\label{fig:OmegaiSPLSA}
\end{figure}

Figure \ref{fig:PredPSDSPLSA} indicates that the Strouhal number of the saddle is not substantially modified by variations in $\delta$. However, as shown in figure \ref{fig:OmegaiSPLSA}, the growth rate of the absolute instability is severely affected by that parameter. For the first Mach number and $k_{sh}=k_{sh1}$, saddles are found for all values of $\delta$, and $\omega_{0i}$ decreases with increasing shear-layer thickness (which also occurs for all other cases). However, for $M_j=1.12$, no saddle is observed for $\delta=0.15$, as the KH mode crosses the acoustic branch before the guided jet mode becomes cut-on. By increasing the shear-layer thickness, that cut-on frequency decreases, and the saddle is recovered. A similar trend on the effect of the shear layer thickness in these waves was also observed by \cite{tam_ahuja_1990}, and more clearly by \cite{mancinelli2020} (who also used a finite thickness model), and numerically by \cite{bogey_gojon_2017}. This suggests that this Mach number is around the end of the A1 branch, and that further increase in $M_j$ will lead to the cessation of that tone. This is confirmed by the analysis of $M_j=1.16, 1.22$, where no A1 saddle is observed for all $\delta$. If $k_{sh}$ is chosen as $k_{sh2}$ (A2 mode), analysis of the eigenspectrum shows that, for $M_j=1.08$, the KH mode never approaches the upstream branch sufficiently to allow for an interaction -- for this case, $k_{sh2}$ is too high to allow for an interaction. That changes for $M_j \geq 1.12$, where saddles are found for all values of $\delta$, and the same trend of decreasing $\omega_{0i}$ with increasing $\delta$ is observed. It is also shown that the spatio-temporal growth rate ($\omega_{i0}$) of A1 modes increases rapidly with increasing $M_j$ (note that $A_{sh}$ is kept constant), which agrees with the trend of increasing amplitude of the screech tone with increasing Mach number in the experiments, suggesting that there may be a connection between $\omega_{i0}$ and the final amplitude of the resonant mode. The same is observed for the A2 modes, but the variation in $\omega_{0i}$ is less steep than for the A1 modes. Interestingly, the model predicts an absolute instability for both A1 and A2 modes for $M_j=1.12$, as in the experiments. For this Mach number, the growth rates of the A1 mode are much greater than the A2 (see also figure \ref{fig:SaddlesA1A2}), suggesting that this mode may be dominant in experiments. This is also in line with the trend observed in the acoustic spectrum, where the A1 peak is greater than the A2.

Thus, the spatially periodic analysis provides a clear explanation for several features of the screech tone, including the appearance of the A1 tone (based on the distance between KH and guided jet modes in the periodic spectrum) and its cut-off behaviour for increasing $M_j$. Consideration of the suboptimal shock-cell wavenumber in the analysis allows for the identification of A2 tones; by comparing the growth rates associated with the saddle-points of both tones, we are able to predict which tone will be dominant for each Mach number. The model also explains why the tone amplitudes increase with $M_j$. Even though all these features are well captured by the model, it also predicts highly amplified A2 tones for $M_j=1.22$, which has a low amplitude axisymmetric peak. Again, this may be due to characteristics of the facility, or the fact that the jet is not exactly a spatially periodic system.

\section{Conclusions}
\label{sec:concl}

The present manuscript proposes a closure mechanism for the A1 and A2 screech modes based of the most energetic wavenumbers associated with the shock-cell structure. Two models are used to evaluate the hypotheses: the first is based on the different waves supported by the flow, and the interaction between wavepackets and the shock-cell structure, considered here as a wave in the mean flow \citep{TamTanna1982,ShenTam2002}. Considering that the energy is extracted from the mean flow mainly by the Kelvin-Helmholtz mode in this framework due to the instability mechanism, the interaction between the wavepacket and the shocks will lead to a redistribution of energy, and other wavenumbers will be energised. At frequencies where the wavenumbers of upstream-travelling waves supported by the jet coincide with this interaction wavenumber, a necessary condition for resonance is achieved, and the jet may screech. It is worth highlighting that this is not a sufficient condition for screech, and that reflection/receptivity processes are not accounted for in the model. Still, screech tones can be predicted fairly well, which suggests that other necessary conditions may be identically satisfied in the jet. Results of the screech prediction method using the first and second peak wavenumbers of the shock-cell structure are in good agreement with experiments for a range of streamwise positions used in the locally parallel analysis for both A1 and A2 modes, in the frequency range of dominance of each mode.

The second model considers a system linearised around a periodic mean flow. Eigenmodes related to the different waves supported by the system are obtained using the Floquet ansatz; due to the periodicity of the spectrum, upstream- and downstream-travelling modes are found in the same region of the spectrum, allowing for interaction. Unstable saddle-points ($\omega_{0i}>0$) involving the KH and the guided jet modes are observed for all values of $M_j$ studied herein, at Strouhal numbers very close to where tones are found in the acoustic spectrum, supporting that screech is caused by this absolute instability mechanism. The model is capable of predicting the cut-on and cut-off behaviour of both A1 and A2 modes (computed using the optimal and first sub-optimal shock-cell wavenumbers, respectively), and also provides a reasoning for the dominance of either A1 or A2 where both modes are supported.

These results support the hypothesis that the interaction between wavepackets and shocks in the ``weakest link'' framework, formulated by \cite{TAM1986}, is a crucial part of the resonance mechanism. In fact, the conditions in which the cited framework were derived are actually very similar to the conditions for the occurrence of a saddle point, as highlighted in equation \ref{eqn:UshocksLinSPLSA}, indicating that the physical argument of energy transfer is actually achieved by means of an absolute instability in the flow. The agreement of both models with experimental data for both A1 and A2 modes suggests that the A2-screech phenomenon occurs as a consequence of the spatial variation of the wavenumber of the shock-cell structure, a result that has not been shown before. This also suggests why theoretical models for the prediction of the A1-A2 mode staging have generally found little success; most models for jet screech do not consider the effect of spatial variation in shock structure; in \cite{mancinelli2019} this is considered inadvertently by different choices of the model parameters, but here it is considered directly from the spatial spectrum of the shock-cell structure. In this framework, dominance of either A1 and A2 modes can be explained by the relative growth rates of each mode at the predicted screech frequency. These conclusions are of relevance not only in providing understanding on the screech phenomenon, but can also be used to design models for screech prediction in other flow configurations. \\

Acknowledgements: This work was supported by the Australian Research Council through the Discovery Project scheme: DP190102220. M.M. acknowledges the support of Centre National d`\'Etudes Spatiales (CNES) under a post-doctoral grant. \\

Declaration of Interests. The authors report no conflict of interest.

\bibliographystyle{jfm}
\bibliography{ref.bib}

\end{document}